\documentclass{emulateapj}

\shorttitle{Discovery of HCNO}
\shortauthors{Marcelino et al.}

\begin{document}

\title{Discovery of Fulminic Acid, HCNO, in Dark Clouds\altaffilmark{1}}

\author{N\'uria Marcelino\altaffilmark{2}, 
Jos\'e Cernicharo\altaffilmark{2}, 
Bel\'en Tercero\altaffilmark{2}, and
Evelyne Roueff\altaffilmark{3}}

\altaffiltext{1}{This work was based on observations carried out with the 
IRAM 30-meter telescope. IRAM is supported by INSU/CNRS (France), 
MPG (Germany) and IGN (Spain)}

\altaffiltext{2}{DAMIR, Instituto de Estructura de la Materia, CSIC, Serrano 121, 28006
Madrid, Spain; nuria@damir.iem.csic.es, cerni@damir.iem.csic.es, belen@damir.iem.csic.es.}

\altaffiltext{3}{Observatoire de Paris-Meudon, LUTH UMR 8102, 5 Place Jules Janssen,
F-92195 Meudon Cedex, France; evelyne.roueff@obspm.fr.}

\begin{abstract}
We report on the first detection in space of fulminic acid, HCNO.
This isomer of HNCO has been observed in three starless cores, B1,
L1544 and L183, and in the low mass star forming region L1527 with a
measured abundance ratio of HNCO/HCNO between 40-70.  However, HCNO
was not detected towards the direction of the cyanopolyyne peak of
TMC-1 or towards the Orion Hot Core region. The derived HNCO/HCNO
abundance ratio in these cases is greater than 350 and 1000 in TMC-1
and Orion, respectively. 
We find that CH$_2$ + NO $\rightarrow$ HCNO + H is a key reaction for
the formation of fulminic acid. A value of 5.5$\times$10$^{-12}$ cm$^3$\,s$^{-1}$
of the corresponding reaction rate coefficient, as given by \citet{Mil03},
allows to reproduce the observed abundances of
fulminic acid in both the observed dark clouds and low mass star
forming core, where the determined abundance of HNCO in these regions
with respect to molecular hydrogen is 1-5$\times$10$^{-10}$.
\end{abstract}


\section{Introduction}
Isocyanic acid (HNCO) was first detected towards Sgr B2 \citep{Sny72}
and its rotational transitions have been found to be prominent towards
the molecular clouds of the Galactic Center \citep{Kua96,Cum86,Tur91,Mart08}.
Since then, it has been observed towards dense cores associated with 
massive star formation \citep{Jac84,Chur86,Zin00}, in the direction of
TMC-1 \citep{Bro81,Kai04}, in diffuse clouds \citep{Tur99} and in
external galaxies \citep{Mei05,Mart06}.
Hence, isocyanic acid is found in a large variety of physical environments. 
In the 3\,mm line survey we have performed in four dark cloud cores using 
the IRAM 30-m telescope \citep{Mar07a,Mar07b} we detected HNCO in all the 
observed sources, showing that this species is a common constituent of dark clouds.
While its abundance in dark and diffuse clouds is $\simeq$10$^{-10}$ \citep{Tur91}, 
in hot cores, the Galactic Center clouds and in external galaxies it is 
$\simeq1-5\times10^{-9}$ \citep{Zin00,Mart06,Mart08,Mei05}.

HNCO has several isomers with a singlet electronic state: HOCN, HCNO, HONC,
and HNOC.
The chemical pathways leading to those isomers could be very different,
since there is not an obvious common precursor. Thus, their detection in 
space could provide some information on the chemistry of these species. 
Unfortunately, only HNCO and HCNO have been fully characterized in the 
spectroscopic laboratories so far (see below).
In this Letter we report on the first detection of fulminic acid, HCNO, 
towards B1, L1544, L183 and L1527.

\section{Observations and Results}
The observations were performed using the IRAM 30-m telescope (Granada,
Spain) between 2002 December and 2007 February, and in 2008 July.
We used two 3\,mm SIS receivers working simultaneously at the same
frequency, but with orthogonal polarizations. Both receivers were tuned
in single sideband mode with image rejections $\sim$24 dB. System
temperatures were $\sim$150 K, for the lowest frequencies and between
200--250 K for the highest ones. 
The observations were done in frequency switching mode.
The spectrometer was an autocorrelator with 40 kHz of spectral resolution 
($\sim$0.13 km\,s$^{-1}$).
Intensity calibration was performed using two absorbers at different temperatures. 
The atmospheric opacity was obtained from the measurement of the sky emissivity 
and the use of the ATM code \citep{Cer85}. Weather conditions were typically 
average summer conditions (water vapor column $\sim$5\,mm and zenith opacities 
$\leq$0.1 at 3\,mm), except for the period in 2008 July when we had very good weather
conditions ($<$2\,mm of precipitable water and opacities $\sim$0.03).

\begin{figure}
\includegraphics[scale=.56]{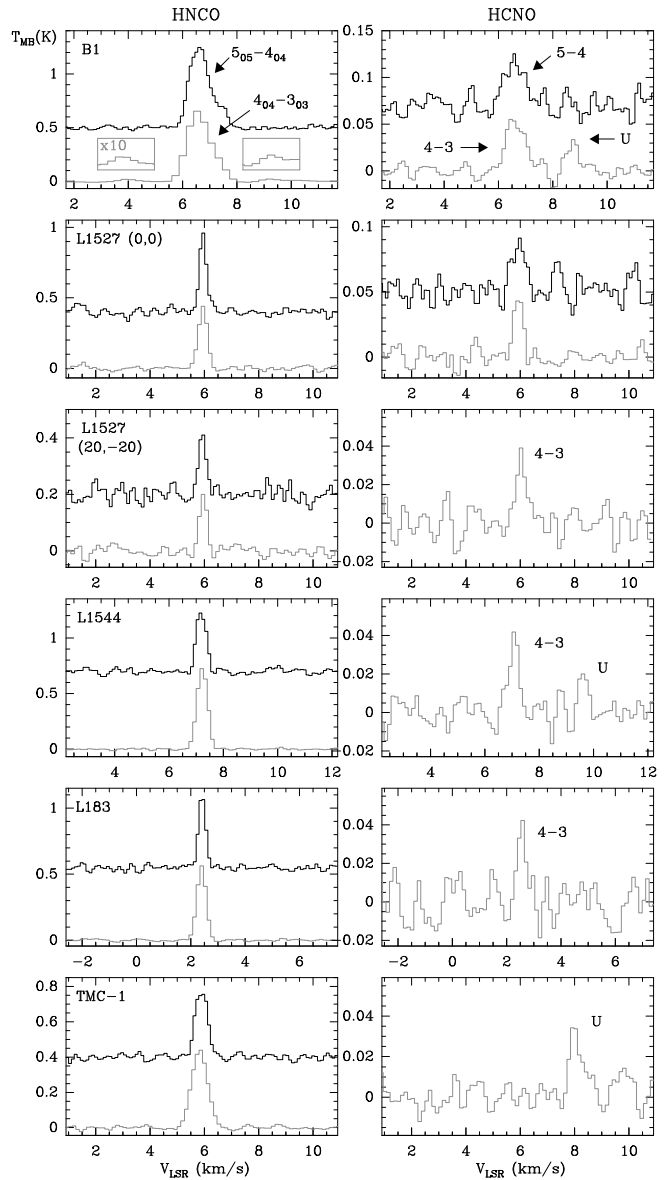}
\caption{Line profiles for the observed transitions of HNCO (left panel),
and HCNO (right panel). Grey lines represent the HNCO $(4_{04}-3_{03})$
and HCNO $(4-3)$ lines, while the black ones are for the HNCO $(5_{05}-4_{04})$
and HCNO $(5-4)$ transitions. 
In B1, the hyperfine components of HNCO $(4_{04}-3_{03})$ at 4 and 9 km\,s$^{-1}$ are shown.
Note the non detection of HCNO $(4-3)$ towards 
TMC-1, and the presence of a unidentified line in this source, L1544 and
B1 at 91750.68$\pm$0.03 MHz. 
\label{fig:fig_hnco-hcno}}
\end{figure}

Pointing and focus were checked, on strong and nearby sources, every 1.5 and 3 hours
respectively.
At the observed frequencies the beamwidth of the antenna is in the range $26''-22''$
and the main beam efficiency is $0.77-0.74$.
The sources being similar in size, or slightly larger, than the main beam of 
the telescope, the contribution of the error beam to the observed intensities
is negligible and all the spectra have been calibrated in main beam temperature scale.
The observed lines are shown in Figure~\ref{fig:fig_hnco-hcno} and 
the derived line parameters, obtained from gaussian fits using the GILDAS
package\footnote{http://www.iram.fr/IRAMFR/GILDAS},
are resumed in Table~\ref{tab:lines}.

\begin{deluxetable}{lcccc}
\tablecaption{HNCO and HCNO observed line parameters\label{tab:lines}}
\tablecolumns{5}
\tablehead{
  Species & $\int T_{\rm MB} dv$ & $V_{\rm LSR}$ & $\Delta v$    &
  $T_{\rm MB}$ \\
   &  (K km s$^{-1}$)     & (km s$^{-1}$) & (km s$^{-1}$) & (K)   \\}
\startdata
\multicolumn{5}{c}{\bf Barnard 1}\\
HCNO \\
$(4-3)$ & 0.065 ( 4) & 6.635 (25) & 1.003 (57) & 0.061 ( 6) \\
$(5-4)$ & 0.064 ( 5) & 6.604 (43) & 1.053 (92) & 0.057 (10) \\
HNCO\\
$(4_{04}-3_{03})$ & 0.722 ( 6) & 6.608 ( 4) & 1.060 (10) & 0.640 ( 5) \\
$(5_{05}-4_{04})$ & 0.728 ( 7) & 6.628 ( 5) & 0.934 (12) & 0.732 (15) \\
\hline
\multicolumn{5}{c}{\bf L1527 (0,\,0)}\\
HCNO\\
$(4-3)$ & 0.020 ( 2) & 5.924 (20) & 0.387 (40) & 0.049 ( 6) \\
$(5-4)$ & 0.023 ( 2) & 5.942 (31) & 0.564 (60) & 0.038 ( 7) \\
HNCO\\
$(4_{04}-3_{03})$ & 0.180 ( 4) & 5.942 ( 4) & 0.386 (10) & 0.439 (12) \\
$(5_{05}-4_{04})$ & 0.181 ( 7) & 5.945 ( 5) & 0.304 (13) & 0.559 (24) \\
\hline
\multicolumn{5}{c}{\bf L1527 (20$''$,\,--20$''$)}\\
HCNO \\
$(4-3)$ & 0.016 ( 3) & 6.036 (33) & 0.450 (92) & 0.034 ( 7) \\
HNCO\\
$(4_{04}-3_{03})$ & 0.068 ( 4) & 5.955 (10) & 0.310 (21) & 0.208 (15) \\
$(5_{05}-4_{04})$ & 0.076 ( 7) & 5.926 (14) & 0.329 (34) & 0.217 (24) \\
\hline
\multicolumn{5}{c}{\bf L1544}\\
HCNO\\
$(4-3)$ & 0.018 ( 2) & 7.073 (24) & 0.420 (64) & 0.040 ( 6) \\
HNCO\\
$(4_{04}-3_{03})$ & 0.359 ( 2) & 7.220 ( 1) & 0.455 ( 3) & 0.741 ( 6) \\
$(5_{05}-4_{04})$ & 0.247 ( 6) & 7.180 ( 5) & 0.419 (11) & 0.554 (21) \\
\hline
\multicolumn{5}{c}{\bf L183}\\
HCNO\\
$(4-3)$ & 0.016 ( 3) & 2.554 (34) & 0.356 (85) & 0.041 ( 9) \\
HNCO\\
$(4_{04}-3_{03})$ & 0.239 ( 2) & 2.393 ( 2) & 0.399 ( 4) & 0.564 ( 6) \\
$(5_{05}-4_{04})$ & 0.183 ( 5) & 2.414 ( 4) & 0.307 ( 9) & 0.561 (17) \\
\hline
\multicolumn{5}{c}{\bf TMC-1 (CP)}\\
HNCO\\
$(4_{04}-3_{03})$ &  0.322 ( 3) & 5.824 ( 4) & 0.687 ( 8) & 0.441 ( 8) \\
$(5_{05}-4_{04})$ &  0.217 ( 5) & 5.889 ( 7) & 0.531 (15) & 0.384 (16) \\
\enddata
\tablecomments{Number in parentheses are 1$\sigma$ uncertainties
in units of the last digits. 
Adopted rest frequencies (MHz) for the observed
lines are 91751.320$\pm$0.004 \& 114688.383$\pm$0.005 for HCNO $J=4-3$ and $5-4$ respectively,
and 87925.240$\pm$0.030 \& 109905.757$\pm$0.030 for HNCO $4_{04}-3_{03}$ and $5_{05}-4_{04}$
respectively.}
\end{deluxetable}

During the 3\,mm line survey (85.9--93.1 GHz) of the dark clouds B1, L1544, L183 and TMC-1,
we have detected several unidentified lines.
Among them one line at 91.751 GHz was observed toward all sources
except TMC-1.
The observed frequency was not found in the available molecular
catalogs: JPL \citep{Pic98}, CDMS \citep{Mul01,Mul05}, and NIST \citep{Lov92,Lov04}.
Using a personal spectral line catalog developed by one of us (J. Cernicharo),
we tentatively identified this line as the transition $J=4-3$ of
fulminic acid (HCNO). The frequencies of this species have been
measured in the laboratory up to $J_{up}=12$, corresponding to $\nu_{max}=275.2$ 
GHz, by \citet{Win71}.
The agreement between the frequency measured in the laboratory
for the $J=4-3$ transition and that observed in dark clouds was better
than 30 kHz, i.e., similar to the uncertainty of the
laboratory measurements and of the velocity determination in these
objects.
Within the survey frequency range we have also observed 
the $4_{04}-3_{03}$ transition of HNCO in all sources, including TMC-1.
Frequencies for HNCO have been computed from the effective rotational
constants obtained recently by \citet{Lap07}.

In order to confirm the new detection we had additional
observations with the IRAM 30-m radio telescope in 2008 July.
We improved the S/N
ratio on the $J=4-3$ transition, and observed the $J=5-4$ towards the
sources where the previous one was detected.
We also included in this run L1527 to our targets.
This source hosts a Class 0/I protostar and lately has been of
great interest because of its strong lines of carbon-chain
species \citep{Sak07,Sak08}.
In this source we have observed two positions, the central one towards the
direction of the protostar, and another one
located at an offset of $(20'',-20'')$, where the emission of carbon chains
is still intense \citep[see,][]{Sak08}. The observation of these two positions
in L1527 could permit to distinguish between the emission in the region around
the protostar and that of the cold envelope.
We have also observed in this run another transition arising from HNCO,
in order to derive volume densities, column densities, and
the abundance ratio between both isomers.

Figure~\ref{fig:fig_hnco-hcno} shows the observed profiles for the HNCO and
HCNO transitions.
The $J=4-3$ transition of HCNO is clearly detected in
B1, L1544, L183 and L1527 well above 6$\sigma$ level.
However, it was not detected towards the direction of the cyanopolyyne peak
of TMC-1, with an rms of 6 mK.
The $J=5-4$ transition was detected only towards B1 and L1527 with similar 
intensities than those of the $4-3$ line.
Upper limits (3$\sigma$) to the $J=5-4$ line intensity are 
$<$0.010 K, $<$0.012 K, and $<$0.018 K for L1527 $(20'',-20'')$, L1544, and 
L183 respectively.
This line should be 2-2.5 times weaker than the $J=4-3$ transition, which is 
consistent with the intensity ratios for HNCO in these sources.

\begin{deluxetable}{lcccc}
\tablecaption{Derived column densities for HCNO and HNCO.
\label{tab:densities}}
\tablehead{
Source    & $N$ (HCNO) &     $N$ (HNCO)           & $n$(H$_2$) & $R$\\
          & $10^{10}$ cm$^{-2}$ & $10^{12}$ cm$^{-2}$          & 10$^5$ cm$^{-3}$ &\\
              }
\tablecolumns{4}
\startdata
B1            &  21(1)/17(4) &  9(1)/8(2)     &  6(1)   &  42/47\\
L1527         &  6(1)/4(1)   &  2.2(6)/1.8(4) &  5(1)   &  37/40\\
L1527-B       &  5(1)/4(1)   &  0.9(5)/0.8(2) &  5(1)   &  18/20\\
L1544         &  8(3)/6(2)   &  5(1)/4(1)     &  0.7(2) &  62/66\\
L183          &  6(2)/5(1)   &  3.4(3)/2.5(6) &  0.5(1) &  57/50\\
TMC-1         &$\leq$1.3/$\leq$1.4       &  5.7(4)/4(1)   &  0.3(1) & $>$390/320\\
Orion-KL      &  $\leq$890   &  9000(2000)    &         & $>$1000 \\
\enddata
\tablecomments{
$R$ is the HNCO/HCNO abundance ratio.
Errors to the column density are in parentheses in units of the last digit. 
L1527-B corresponds to the position $(20'',-20'')$ of this source. 
The first entry in $N$(HNCO), $N$(HCNO), and $R$ corresponds to the values
obtained from the rotational diagrams, while the second one corresponds to those 
obtained from the $LVG$ calculations.
}
\end{deluxetable}

\section{Discussion}
The dipole moment of fulminic acid is 3.099 D \citep{Tas89},
while that of HNCO is 2.1 \citep[$\mu_a$=1.60 D, $\mu_b$=1.35 D;][]{Hoc74}. 
Consequently, both molecules could have different excitation
conditions and
we have performed rotational diagrams in order to obtain rotational
temperatures ($T_{\rm rot}$) and column densities ($N$). 
Two transitions of HNCO were observed in all the sources.
The fit of both lines gives rotational temperatures of $12\pm3$ K, $12\pm4$ K,
$15\pm6$ K, $6\pm1$ K, $7\pm1$ K and $6\pm1$ K for B1, L1527 $(0,0)$, L1527 
$(20'',-20'')$, L1544, L183 and TMC-1, respectively.
Errors to the derived temperatures and column densities include those obtained from 
the gaussian fit, calibration errors and the fact that we are using only two transitions 
to perform the rotational diagram. 
Although HNCO presents hyperfine structure, we have not
included it in our calculations since it is only important for
low rotational quantum numbers ($J<3$). Nevertheless, two weak hyperfine lines are
visible towards B1 in the 4$_{04}$-3$_{03}$ transition at 4 and 9 km\,s$^{-1}$
(see Fig.~\ref{fig:fig_hnco-hcno}).
They are ten times weaker than the strongest
component indicating that the HNCO lines are optically thin.
B1 and the central position of L1527 are the only sources where two
transitions of fulminic acid have been detected, and where we have
estimated rotational temperatures of $10\pm4$ K and $17\pm8$ K respectively. 
For the other three sources $T_{\rm rot}$ has been assumed to be the same for both isomers.
Table~\ref{tab:densities} shows the obtained column densities for HCNO and
HNCO, and the abundance ratio between both isomers.

Collisional rates for HNCO are available from Green's calculations
\footnote{unpublished results, see http://molscat.giss.nasa.gov}.
We have used a Large Velocity Gradient code to derive volume
and column densities for HNCO and HCNO in order to better constraint the
HNCO/HCNO abundance ratio (hereafter refered to as $R$).
The collisional rates $J\rightarrow J'$ for HCNO
are unknown but we have assumed that they are identical to 
the $J_{0J} \rightarrow J'_{0J'}$ collisional rates of HNCO.
The adopted kinetic temperatures
are 10 K for all dark clouds except B1 and L1527 for which we have assumed
$T_{\rm kin}=15$ K following the rotational temperatures derived above from HNCO.
The obtained values for $n$(H$_2$), $N$(HNCO) and $N$(HCNO) are
also given in Table~\ref{tab:densities}. 
Errors to the $LVG$ calculations are estimated to be about 25 \%.
The agreement between the column
densities derived from the two methods is very good.

In Orion-KL we have detected more than 60 spectral lines of HNCO  
at 3, 2 and 1.3\,mm (see the line survey of Orion by Tercero et al., 2008, 
in preparation).
Column densities have been calculated using LTE approximation included
in radiative transfer codes developed by J. Cernicharo (Cernicharo
2008, in preparation). We have kept constant the physical properties
of each component of Orion-KL (extended ridge, compact ridge, plateau
and hot core; see \citealt{bla87} for the standard values we have
assumed) including their sizes and offsets with respect IRc2 (survey
pointing position). Corrections for beam dilution are applied for each
line. We have estimated the uncertainty to be about 25 \%, considering 
error sources as the line
opacity effects, the simplification of the model, the division of the
whole line emission in different components and pointing and
calibration errors. The results are
(1.5$\pm$0.4)$\times$10$^{14}$, (1.5$\pm$0.4)$\times$10$^{15}$, 
(1.0$\pm$0.3)$\times$10$^{15}$ and (7$\pm$2)$\times$10$^{14}$
cm$^{-2}$ for the extended ridge, the plateau, the compact ridge and
the hot core respectively. For the hot core a second component with
$T_{\rm kin}$=300$\pm$75 K and $N$(HNCO)=(6$\pm$2)$\times$10$^{15}$
cm$^{-2}$ is needed
to reproduce the observations. In none of these components HCNO has
been detected. The upper limit to $R$ is $\>$1100 in the hot core, the
plateau and the compact ridge, and $\>$300 in the extended ridge.
Since the Orion region is dominated by grain-surface chemistry
--which explains the large column density of HNCO--
rather than ion-molecule and neutral-neutral gas phase reactions, the 
lack of HCNO suggests that grain surface chemistry is not the main path
to the production of HCNO in molecular clouds (see below).

From recent state of the art  {\it ab initio} calculations, the relative energies of HOCN, HCNO and HONC, 
in their ground electronic singlet states, compared to HNCO, are 24.7, 70.7 and 84.1 kcal mol$^{-1}$
at an unprecedented level of accuracy \citep{Sch04}. Another isomer, HNOC, is predicted by {\it ab initio} calculations 
at $\simeq$135 kcal mol$^{-1}$ above HNCO \citep{Meb96}.
The relative abundance of the different isomers in molecular clouds depend on the chemical paths leading 
to their formation. Unlike the case of HCN/HNC, which are mainly formed from the dissociative recombination 
of HCNH$^+$, there is not an obvious common parent molecule leading to the simultaneous formation of the 
various HNCO isomers. 
The most evident formation path for HNCO and HCNO could be the reaction of HCNH$^+$ 
with atomic oxygen and OH. Unfortunately, 
it has been shown in the laboratory that HCNH$^+$ does not react with oxygen \citep{Sco99}
and nothing is known about its reaction with OH.
\citet{Igl77} suggested the reaction NCO$^+$  + H$_2$ $\rightarrow$ HNCO$^+$ + H
as a starting point for the chemistry of HNCO.
The other isomers can not be produced via this reaction as the corresponding channels are endothermic.
We have introduced a tentative gas phase chemical network with the three most stable isomers --HNCO, HOCN and HCNO--
and their associated ions which have been studied theoretically by \citet{Lun96}.
We have obtained that the proton transfer reactions involving H$_3^+$ and NCO, the most stable isomer, could lead 
to both HCNO$^+$ and HOCN$^+$ via exothermic channels. The HNCO$^+$, HCNO$^+$ and HOCN$^+$ ions can 
further react with molecular hydrogen, giving respectively H$_2$NCO$^+$, H$_2$CNO$^+$ and HOCNH$^+$, 
whose relative energies have been computed by \citet{Hop89}. These ion-molecule reactions are followed 
by dissociative recombination where we assume that no chemical structural change occurs so that 
 H$_2$NCO$^+$/H$_2$CNO$^+$  + e$^- \rightarrow$ HNCO/HCNO + H,
 and
 HOCNH$^+$  + e$^- \rightarrow$ HNCO/HOCN + H.

\begin{deluxetable}{lcccccc}
\tablecaption{Model abundances obtained for the HNCO isomers in cold clouds
\label{tab:model_abundances}}
\tablehead{
$n$(H$_2$)     &NO & HNCO & HCNO & HOCN & NCO & $R$ \\
10$^4$ cm$^{-3}$ &  10$^{-7}$ & 10$^{-10}$  & 10$^{-11}$  & 10$^{-13}$  &  10$^{-8}$   \\}
\tablecolumns{6}
\startdata

0.1  & 0.005 & 0.004 & 0.04  & 0.47  & 0.02 & 0.8 \\
0.3  & 0.01  & 0.01  & 0.08  & 1.30  & 0.07 & 1.5 \\
1.0  & 8.6   & 18    & 6.30  & 38.0  & 6.6  & 29  \\
3.0  & 7.4   & 18    & 3.10  & 12.0  & 4.5  & 58  \\
10   & 5.1   & 9.3   & 1.20  & 2.0   & 2.0  & 72  \\
30   & 3.2   & 3.4   & 0.58  & 0.31  & 0.8  & 58  \\
50   & 2.5   & 2.0   & 0.41  & 0.13  & 0.5  & 49  \\
100  & 1.7   & 0.9   & 0.25  & 0.04  & 0.3  & 36  \\
\enddata
\tablecomments{$R$ is the HNCO/HCNO abundance ratio}
\end{deluxetable}

Neutral-neutral reactions may also be at work. So, HNCO can be formed by 
CN + O$_2$ $\rightarrow$ NCO + O, followed by NCO + H$_2$ $\rightarrow$ HNCO + H,
as suggested by \citet{Tur00}. However the presence of an activation barrier in the
NCO + H$_2$ reaction can not be dismissed and a tentative barrier of 500 K is 
introduced in the present work. In addition, a favorable neutral-neutral pathway 
leading to the formation of HCNO is provided by
CH$_2$  + NO $\rightarrow$ HCNO + H, a reaction well studied in combustion and 
atmospheric chemistry \citep{Gla98}, for which no activation barrier has been 
found by \citet{Rog98} from {\it ab initio} studies.
Other neutral-neutral reactions of HCNO in the presence of atomic oxygen and NO 
may occur \citep{Mil03}. Including these various possibilities in
a chemical network, we have
computed equilibrium solutions by solving directly the steady state
chemical equations.
Table~\ref{tab:model_abundances} displays the fractional abundances
relative to molecular hydrogen for different volume densities in cold clouds,
at a temperature of 10 K and a cosmic ionization rate of 5$\times10^{-17}$ s.
These results do not differ considerably when assuming a temperature of 15 K and 20 K.
The predicted abundance for NO is in good agreement with the value observed
in dark clouds \citep{Ger92,Ger93}.
NCO has also a large abundance, but due to its moderate dipole moment (0.67 D) and a $^2\Pi$
structure diluting the emission spectrum resulting from an enhanced partition function,
the expected intensity from its rotational lines is rather weak. From our line
survey of dark clouds \citep{Mar07a,Mar07b} we derive limits to its
abundance of 1-5$\times10^{-9}$.
The result from our models show a good agreement with the values observed for $R$ in dark clouds
and low-mass star forming regions. 
It increases with volume density up to $n$(H$_2$)=10$^5$ cm$^{-3}$, but decreases again when the density is 
larger than this value. It seems that for low and high volume density clouds fulminic acid could be rather 
abundant, while in clouds of moderate density, like TMC-1, it will be much less abundant than isocyanic acid, HNCO.
Furthermore TMC-1 shows a very peculiar chemistry, being a carbon rich source while it is deficient in oxygen bearing 
species. That could be the reason why HCNO is detected in L183, similar to TMC-1 in density and temperature but oxygen 
rich, and not towards TMC-1.
The formation of HNCO on dust grains may also be very efficient \citep{Has93} and these surface chemistry 
processes could also lead to the formation of HCNO (Herbst, private communication). However, the gas phase 
models predict abundances for HNCO very close to those observed in dark clouds. Moreover, these models also 
predict the observed HNCO/HCNO abundance ratio. 
The detection of the different isomers of HNCO provides a fundamental information concerning the physical and
chemical conditions, in particular the respective contribution of gas phase and grain processes in the chemistry
since there is not an obvious common precursor for them. Thus,
the observation of the other isomer of HNCO, HOCN, for which present
gas phase models predict a very low abundance in dark clouds could provide a good discrimination between gas and grain
surface processes in dark clouds.
The search for HNCO isomers towards different environments would be the next step to further investigate their
chemical production.

\acknowledgments
We would like to thank Eric Herbst for useful comments and suggestions, 
and Marcelino Ag\'undez for carefully reading of the paper.
We also thank our anonymous referee, whose detailed comments have helped to
improve the paper.
This work has been supported by Spanish Ministerio de Ciencia e Innovaci\'on 
through grants AYA2003-2785, AYA2006-14876, and by DGU of the Madrid 
community government under IV-PRICIT project S-0505/ESP-0237 (ASTROCAM).

\end{document}